\documentclass[twocolumn,showpacs, preprintnumbers,amsmath,amssymb, superscriptaddress]{revtex4}

\usepackage{graphicx}
\usepackage{dcolumn}
\usepackage{bm}
\usepackage{color}

\begin{document}

\preprint{}

\title{
Transition to diversification by competition for resources in catalytic reaction networks
}

\author{Atsushi Kamimura}
\affiliation{Department of Basic Science, 
The University of Tokyo,
3-8-1, Komaba, Meguro-ku, Tokyo 153-8902, Japan}
\author{Kunihiko Kaneko}
\affiliation{Department of Basic Science, 
The University of Tokyo,
3-8-1, Komaba, Meguro-ku, Tokyo 153-8902, Japan}
\date{\today}

\begin{abstract}
All life, including cells and artificial protocells,  
must integrate diverse molecules into a single unit in order to reproduce.
Despite expected pressure to evolve a simple system with the fastest replication speed,
the mechanism by which the use of the great variety of components, and the coexistence of diverse cell-types with different compositions are achieved is as yet unknown. 
Here we show that coexistence of such diverse compositions and cell-types is 
the result of competitions for a variety of limited resources.
We find that a transition to diversity occurs both in chemical compositions and in protocell types, 
as the resource supply is decreased, when the 
maximum inflow and consumption of resources are balanced.
\end{abstract}


\maketitle
Cells, even in their primitive forms\cite{protocell, Blain, Libchaber},  
must integrate diverse molecules into a single unit so that they keep reproduction.
The use of diverse chemical components is sustained from the origins of life\cite{Dyson}, 
while biodiversity in ecosystems is emphasized as salient characteristics\cite{May, McCann, Loreau} even at a cellular level\cite{Hughes, Whitham}.
How diversity originated in cellular reproduction is an important open question\cite{Dyson}.
One may naively expect that a simple replicator consisting of only a few molecular 
species would evolutionarily triumph as it can reproduce faster than a complex system 
using diverse components. Additionally, as replicators with the fastest division speeds 
would dominate, coexistence of diverse replicators would not be expected. 
In fact, several $in$ $silico$ artificial life models show dominance of such simple replicators\cite{Fontana-Buss,Tierra}. 

In nature, however, diverse components exist within a cell and diversity of cell types(replicators) is sustained\cite{Barabasi, JainKrishna, MehrotraJain}. 
Are both $``compositional"$ diversity at an individual level and $``phenotypic"$ diversity at the population level present at the level of ``protocells", a primitive early stage of cellular evolution, in spite of the selective pressure for survival?

Chemical protocell replicators have been investigated using the hypercycle model, in which different molecular species mutually catalyze the replication of each other\cite{EigenSchuster, Eigen}. 
Protocells encapsulating such hypercycles can exhibit robust reproduction even in the presence of parasitic molecules that may destroy the mutual catalytic reactions \cite{MaynardSmith, Szathmary, McCaskill, Hogeweg, Kaneko, Kaneko2, KamimuraKaneko}. 
 In these studies, it is often assumed that chemical resources for 
 synthesis of biopolymers are supplied sufficiently. 
 Thus, a protocell consisting of a simple hypercycle with few components replicates 
 faster, and quickly dominates the pool.
 
As the catalytic activities of molecules and cell populations increase, however,
nutrient depletion or resource competition is inevitable\cite{Chen}. 
When only a single resource is provided, 
competition for the limited resource leads to 
survival of only the fittest protocell type.
When $multiple$ resources are competed for, then, 
do protocells diversify into distinct types specialized for the use of different resources, 
and is coexistence of diverse cell types possible? 
Here, we show through numerical simulation of a model of interacting protocells consisting 
of hypercycle reaction networks that such a transition to increased diversity 
occurs when available resources are limited.  

\begin{figure}[t]
\begin{center}
\includegraphics[width=7.1cm]{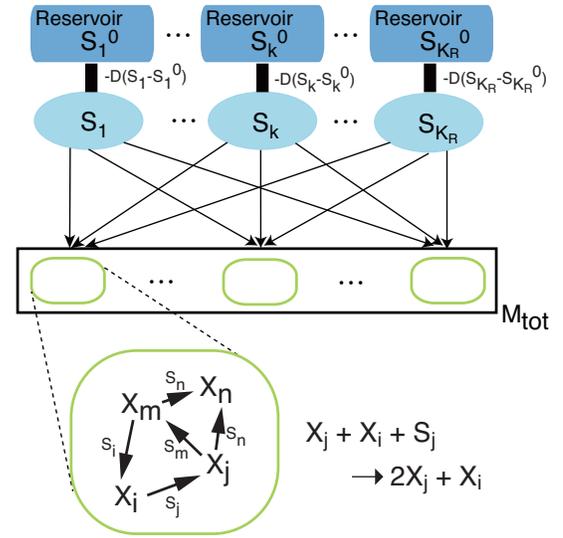}
\vskip 0.25cm
\caption{Schematic representation of our model. 
The system is composed of $M_{\rm tot}$ cells, each of which contains molecule 
species $X_j$ ($j=1,...,K_M$) that form a catalytic reaction network to replicate each $X_j$. The cells share each resource $S_{k}$, which is consumed by the replication of molecule $X_k$. 
The resources flow into a common bath for the cells from an external 
reservoir (environment) via diffusion $-D(S_k-S_k^0)$($k = 1,...,K_R$), where $S_k^0$ is a randomly-fixed constant $S_k^0 \in [0,M_{\rm tot}]$ and $D$ is the diffusion constant.}
\label{fig1}
\end{center}
\end{figure}

For this purpose, we consider a protocell model in which each molecule 
($X_j; $$j=1,..,K_M$) replicates with the catalytic aid of $X_i$, 
predetermined by a random network, by consuming a 
corresponding resource ($S_{j};$$j=1,...,K_R=K_M)$.
The outline of the model is as follows(Fig. \ref{fig1}).
There exist $M_{\rm tot}$ protocells, consisting of $K_M$ species of replicating molecules where some possibly have null population. 
Molecules of each species $X_j$ are replicated with the aid of some other catalytic molecule $X_i$, determined by a random catalytic reaction network, by consuming a predetermined resource $S_{j}$, one of the supplied resource chemicals $S_k$$(k=1,...,K_R)$, as follows:
\[
X_j + X_i + S_{j} \xrightarrow{c_i} 2X_{j} + X_i.
\]
For this reaction to replicate $X_j$, one resource molecule is needed, and 
the replication reaction does not occur $S_{j}<1$. 
When $K_M=K_R$, each resource species $S_j$ corresponds to 
each molecule species $X_j$, while for $K_M > K_R$, 
a common resource molecule species is used for replication of multiple 
molecule species $X_{j_1},X_{j_2}..$, as discussed in the supplement.
The reaction coefficient is given by the catalytic activity $c_i \in [0,1]$ of the molecule species $X_i$.
With each replication, error occurs with probability $\mu$ as shown below.

The resources ($S_j$) diffuse with diffusion constant $D$ through a common medium for a population of protocells from external reservoirs of concentrations $S_j^0$. 
$D$ controls the degree of the resource competition, as the resource supply is limited with decreasing $D$.

For each molecule species, the density for the path of the catalytic reaction is 
given by $\rho$(which is fixed at 0.1) so that each species has $\rho K_M$ catalysts on 
average. Once chosen, the reaction network is fixed throughout each run of simulations. 
Autocatalytic reactions in which $X_i$ catalyzes the replication of itself are excluded
 from the random catalytic network, and direct mutual connections are excluded so that $X_j$ does not work as a catalyst for $X_i$ if $X_i$ is the catalyst for $X_j$.

Structural changes may occur during replication to alter monomer sequences of 
polymers and catalytic properties of the molecule. 
In the present model, this alteration is included as a random change to other molecular species during the replication process.
When replication of $X_j$ occurs, it is replaced by another molecule $X_l$ ($l \neq j$) with a probability $\mu$.
For simplicity, we assume that this error leads to all other molecule species with equal probability, $\mu/(K_M-1)$ where $K_M$ is the number of molecule species.

When the total number of molecules in each cell exceeds a given threshold $N$, 
the cell divides into two and randomly partitions molecules, and one randomly chosen 
cell is removed from the system in order to fix the total number of cells at $M_{\rm tot}$.

\begin{figure}[t]
\begin{center}
\includegraphics[width=14cm]{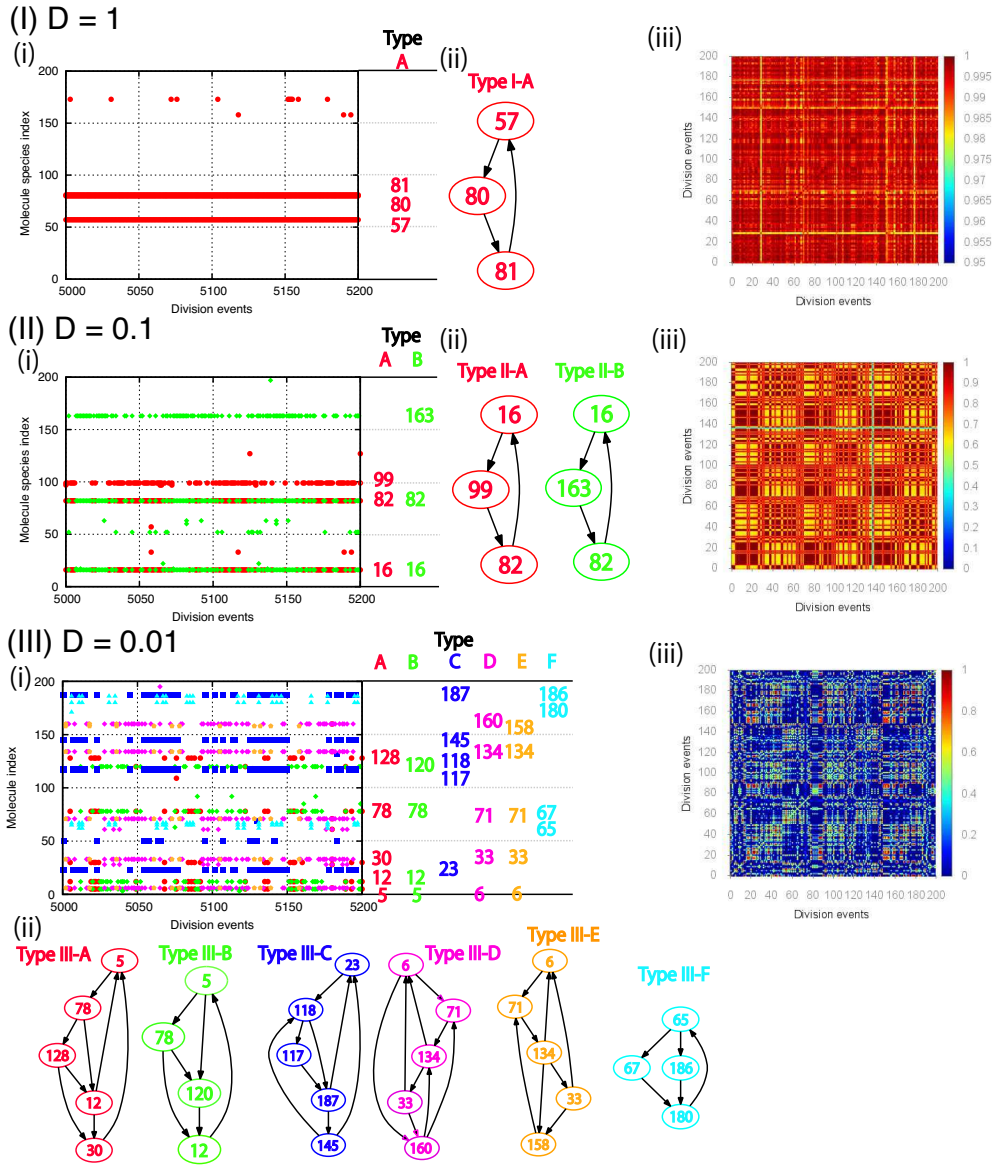}
\vskip 0.25cm
\caption{ Compositions and types of reproducing protocells for 
(I)$D=1$, (II)$D=0.1$, and (III)$D=0.01$. 
Compositions of molecule species are shown in successive 200 division events 
in the system with $K_M = K_R$.
(i) At successive division events (abscissa), molecule species indices with 
more than 20 copy numbers in the dividing cell are marked. 
Cells are categorized into a few, or, several types, indicated by different colors, 
according to the majority set of molecule species. For each type, 
the indices of majority species are shown to the right of the figures. 
(ii)The catalytic network formed by the majority molecule species is shown for each cell type. Each numbered node corresponds to a molecule species, and 
the arrow from index $i$ to $j$ represents catalysis of $X_j$ replication by $X_i$. 
(iii) Similarities $H_{ij}$ between cells $i$ and $j$, as defined in the text, are plotted over 200 successive division events
, using a color code from $H_{ij}\sim 1$(red) to $H_{ij}\sim 0$(dark blue). 
If the similarity is close to one, the composition is almost preserved by cell division, 
while zero similarity indicates that a reproduced cell has completely different composition. Intermediate values of similarity indicate the overlap in some molecule species between the cells, as in Type II-A and -B, Type III-A and -B, and Type III-D and -E.  
Parameters are $K_M = K_R = 200$, $M_{\rm tot} = 100$, $N = 1000$, and $\mu = 0.001$.}
\label{fig2}
\end{center}
\end{figure}

We simulated the model by changing the speed of resource supply $D$. 
When the resources are supplied sufficiently fast (e.g., for $D = 1$), 
a recursively growing state is established with a few molecular species, 
where the composition is robust against noise and perturbations by 
the division process. 
In this state, the (typically) three primary components form a three-component 
hypercycle. Other molecular species in a cell are typically catalyzed by a member 
of the hypercycle.
All the dividing cells adopt this three-component hypercycle, 
thus, there is neither compositional nor phenotypic diversity(Fig. \ref{fig2}(I)).

To check cell reproduction fidelity, we introduced similarity between cells as follows: 
as each cellular state is characterized by the number of molecules of each 
chemical species $\vec{N}_i = (n_1, n_2, ..., n_{K_M})$, 
similarity is defined as the inner product of these composition vectors 
between two cell division events, i.e.,
$H_{ij} = \vec{N}_i \cdot \vec{N}_j / (|\vec{N}_i || \vec{N}_j |)$ 
between the $i$-th and $j$-th division events. 
In the above case, the similarity between mother and daughter cells is close to one, implying high-fidelity reproduction.

As $D$ decreases below 0.1, phenotypic diversity starts to increase. 
For example, two cell types(II-A,B) coexist in Fig. \ref{fig2}(II) and 
consist of three-component hypercycles differing by one component. 
Both types divide with approximately equal speed and coexist over $10^2$ 
generations. 
In 200 successive division events(Fig. \ref{fig2}(II)(iii)), 
one type has a similarity near unity(red), and the similarity of the 
other ranges between 0.6 and 0.7(yellow), 
implying that the two types mostly reproduce themselves with a small 
probability (approximately 0.01/division) of switching types.
Over much longer generations, replication errors can produce different 
types capable of replacing the existing cell types.

As $D$ decreases further, both the phenotypic and compositional diversity
continue to increase.
For $D=0.01$(Fig. \ref{fig2}(III)), six cell types ($A-F$) appear. 
Each type forms a distinct hypercycle network in which the species catalyze replication of each other. 
Here, some types (III$-A$ and $B$, III$-D$ and $E$) share some 
common molecular species, while the others do not.
Similarities are approximately equal to unity(red) for some cells, 
while cell types that have the common chemical components(yellow to light blue) as 
well as cell types with completely orthogonal composition(dark blue) appear from time to time 
(Fig. \ref{fig2}(III)(iii)). Also, the number of replicating chemical species 
in each cell is slightly increased(see Fig. \ref{fig2}(III)(ii)).
As $D$ decreases to the order of $0.001$, 
more cells with lower similarity appear, and compositional and phenotypic diversity further increase.
 
\begin{figure}[t]
\begin{center}
\includegraphics[width=15cm]{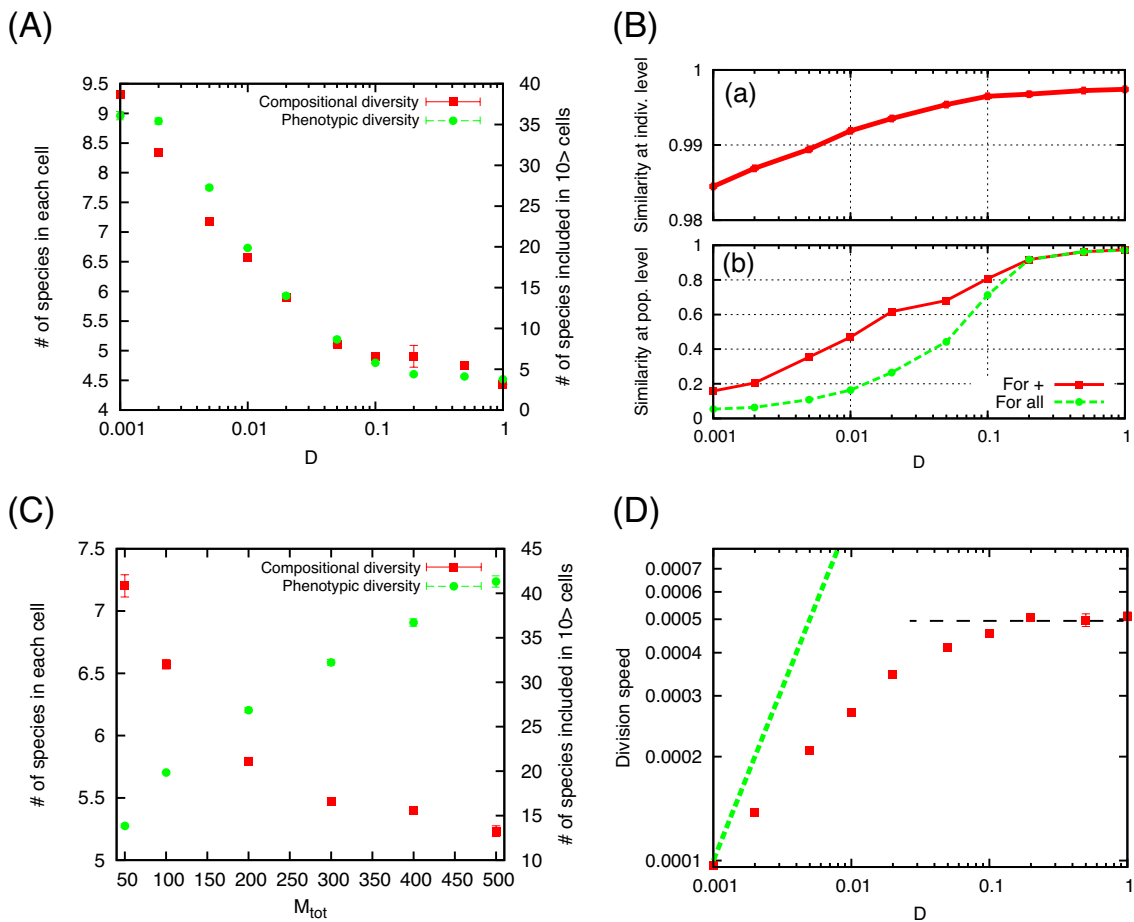}
\caption{(A) Compositional and phenotypic diversity plotted as a function of $D$. For compositional and phenotypic diversity, the numbers of chemical species included in each cell (left ordinate), and in more than 10 cells out of $M_{\rm tot}$ cells (right ordinate) are shown.
(B)(a) Average similarity of mother and daughter cells and (b) average similarity between cells in $10^4$ successive division events in 30 different networks, plotted against $D$. For (b), 
the average over cells with positive similarity(red) is shown in addition to the 
average for all cells (green).
(C) Dependence of compositional (left ordinate) and phenotypic (right ordinate) diversity on the number of cells, $M_{\rm tot}$, with fixed $D = 0.01$. 
(D) Average division speed as a function of $D$. For reference, a linear line with $D$ is plotted. For $D > 0.2$, an estimate of the division speed ($\sim 0.0005$) is also shown with a dotted line (see main text).
Unless otherwise indicated, the data are obtained as the average over $10^5$ division events in 30 different networks. The parameters are $K_M = K_R = 200$, $N=1000$, and $\mu = 0.001$.
}
\label{fig3}
\end{center}
\end{figure}

We statistically studied the quantitative dependence of diversity upon 
the parameter $D$ using a variety of networks. 
As $D$ decreases below $\sim 0.1$, the compositional diversity of 
each protocell and the phenotypic diversity at the population level
increase (Fig. \ref{fig3}(A)).
With the increase in diversity, reproductive fidelity decreases both 
at an individual level (Fig. \ref{fig3}B(a)) and at a population level, i.e., 
over all pairs of $10^4$ division events in 30 runs (Fig. \ref{fig3}B(b)). 
Protocells with different hypercycles start to appear below $D\sim 0.1$(Fig. \ref{fig2}).

The transition to increased diversity generally occurs 
for sufficient resource diversity, i.e., for large $K_R$, 
independent of reaction network choice. 
The phenotypic diversity increases as $\sim K_R$, 
but is bounded by the finite number of interacting cells,$M_{\rm tot}$(Supplementary Information). 
As $M_{\rm tot}$ increases, the number of coexisting cell-types increases, 
while the compositional diversity, i.e., 
the number of components in each cell type, decreases(see Fig. \ref{fig3}(C)).
This trade-off between compositional and phenotypic diversity suggests 
that each cell-type is 
specialized for fewer chemical components as the number of cell types increases.

Altogether, the data show transition behavior around $D = D_c \sim 0.1$.
Below $D_c$, the division speed decreases, while above $D_c$ it is approximately constant (Fig. \ref{fig3}(D)). 
This suggests that at the transition point the maximum inflow and consumption rates of resources are balanced.
The maximum inflow rate is estimated as $D \bar{S_j^0}$, 
where $\bar{S_j^0}$ is a typical reservoir concentration. 
The intrinsic consumption rate by all cells is estimated to be $M_{\rm tot} \bar{c_i} p_j^c$, where $\bar{c_j}$ is a typical catalytic activity, and $p_j^c$ is the probability of picking up a pair that replicates the molecule $X_j$. 
When sufficient resources are available, 
the three-component hypercycle dominates, and $p_j^c \sim 1/9$.
Therefore, the value of $D$ at which the maximum inflow and consumption rates of resources are balanced is estimated as 
$D = M_{\rm tot} \bar{c} /9\bar{S_j^0}$. 
Since $c_i$ is distributed homogeneously in $[0,1]$, 
its average value is 0.5; however the remaining components are typically 
biased to have higher catalytic activities, and $\bar{c}$ therefore typically ranges
from $0.7$ to $0.8$. On the other hand, $S_j^0$ is distributed homogeneously in 
$[0, M_{\rm tot}]$, so the simple average of $\bar{S_j^0}$ is $50$ with $M_{\rm tot} = 100$, but the remaining 
components are biased with have $\bar{S_j^0} \sim 70$. Thus the critical value of $D$ is approximately $0.11-0.13$.
It is noteworthy that below $D_c$ the division speed decreases more slowly than the inflow rate, as indicated by the line proportional to $D$(Fig. \ref{fig3}(D)), 
suggesting cells can utilize more diverse resources for growth by increasing the available number of resource species.

For $D > 0.2$, 
sufficient resources are available, thus, the intrinsic reaction rate of 
the three-component hypercycle is the main determinant of the division speed. 
The probability, $p_c$, for picking up a pair between which a catalytic reaction exists is $\sim 1/3$. Using the above typical catalytic activity $\bar{c}$, the division speed is estimated as $2\bar{c}/3N$, which is approximately $0.0005$ for $N=1000$(See Fig. \ref{fig3}(D)).

Why does the transition to diversity occur with the decrease in resources?
A simple case illustrates this diversity transition from dominance of 
a single type to coexistence of various types. 
Consider two types of cells that compete for resources. 
One type ($A$) consists of molecules $X$ and $Y$ 
while the other type ($B$) consists of molecules $X$ and $Z$. 
Each of molecule $X$, $Y$, and $Z$ is replicated by mutually catalytic reactions utilizing resources, 
$S_X$, $S_Y$, and $S_Z$, respectively.  
When supplied sufficient resources, the population of each cell type 
grows exponentially, where the growth rate 
is proportional to the number $n_{A(B)}$ of each cell type $A$ and $B$, i.e., $dn_{A(B)}/dt \propto \gamma_{A(B)} n_{A(B)}$, and the proportionality coefficient, $\gamma_{A(B)}$, is proportional to resource abundances $S_{Y(Z)}$. In this case, Darwinian selection works so that the type 
with a larger $\gamma$ dominates. 
This selection process works as long as the resources are 
sufficient. 
However, when all the resources are limited, 
competition for the available resources $S_{Y(Z)}$ among $n_{A(B)}$ cells decreases $\gamma_{A(B)}$ so that it is inversely proportional to $n_{A(B)}$, respectively.
The population dynamics are, therefore, represented by 
$dn_{A(B)}/dt \propto c_{A(B)}$, with constant $c_{A(B)}$, when the maximum inflow 
is decreased to balance with the rate of resource consumption. 
In this form, the solution with two coexisting types is known to be stable\cite{Biebraicher, Eigen}(Supplemental Information). 
The transition to diversity in composition and phenotypes in the original model is based on this change from 
exponential to linear growth due to resource limitation.  
On the other hand, the above argument on the diversity transition supports the generality of our result, as the 
number growth in chemical components by catalytic reactions changes from exponential to linear with the decrease in resources.

When life originated, a set of diverse, self-replicating catalytic polymers(replicators) would 
emerge from the primordial chemical mixture, to make a reproducing protocell. 
Although the importance of molecular and protocellular diversity has been noted by 
Dyson\cite{Dyson}, their origins are not well addressed, especially as 
compared with diversity in ecological systems. Our results indicate 
that competition for a variety of limiting resources can be a strong driving force to 
diversify intracellular dynamics of a catalytic reaction network and to develop 
diverse protocell types in a primitive stage of life. 
Indeed, it is natural that diverse chemical resources are available in the 
environment, and competition for resources increases as the protocells 
reproduce and more cells compete. Thus, 
diversification in composition and protocell types is an inevitable outcome. 

According to our results, the diversification is understood 
as a kind of phase transition in population dynamics with decreased resource.
In ecology, 
the niche dimension hypothesis in plant communities was proposed, 
in which the number of coexisting species increases with 
the number of limiting factors increases while greater resource abundance 
decreases diversity\cite{Tilman, Tilman2, Hutchinson}. 
Our results suggest that such population dynamics in ecology are 
possible with primitive mixtures of catalytic 
molecules competing for a variety of resources\cite{Lancet}, even without the need for genetic changes as in speciation.

This work is supported by the Japan Society for the Promotion of Science. This work is also supported in part by the Platform for Dynamic Approaches to Living System from the Ministry of Education, Culture, Sports, Science, and Technology of Japan, and the Dynamical Micro-scale Reaction Environment Project of the Japan Science and Technology Agency.

\newpage

\end{document}


\preprint{}

\title{
Supplementary Material for \\Transition to diversification by competition for resources in a catalytic reaction network
}

\author{Atsushi Kamimura}
 \affiliation{Department of Basic Science, 
The University of Tokyo,
3-8-1, Komaba, Meguro-ku, Tokyo 153-8902, Japan}
\author{Kunihiko Kaneko}
\affiliation{Department of Basic Science, 
The University of Tokyo,
3-8-1, Komaba, Meguro-ku, Tokyo 153-8902, Japan}
\date{\today}

\begin{abstract}

\end{abstract}

\maketitle

\section{A simple case - Mutually catalytic molecules - }

\subsection{Reactions}
Let us consider two types of cells. One type, denoted by $A$, 
consists of molecule species $X$ and $Y$. 
The other type, $B$, consists of molecule species $X$ and $Z$. 
The molecule species mutually catalyze the replication of each other to form a minimal hypercycle as follows:
\[ X + Y + S_X \rightarrow 2X+Y, \hspace{5mm} Y + X+ S_Y \rightarrow 2Y+ X \]
and 
\[ X + Z + S_X \rightarrow 2X+Z, \hspace{5mm} Z + X + S_Z \rightarrow 2Z+ X. \]
We denote the intrinsic catalytic activities of $X$, $Y$, and $Z$, respectively, by $c_X$,  $c_Y$ and $c_Z$.
Each reaction to synthesize $X$, $Y$, and $Z$ utilizes the resource $S_X$, $S_Y$, and $S_Z$, respectively, which are shared
by cells.

The intrinsic reaction rates for replicating the molecule species $X$ are given by
\[ F_X^A = V_A f_X^A s_X, \hspace{5mm} F_X^B = V_B f_X^B s_X, \]
respectively, for cell type $A$ and $B$. 
Here, $V_A$, and $V_B$ are the volumes of cell types $A$ and $B$, 
respectively, and 
$f_X^A = c_Y x^A_X x^A_Y, f_X^B = c_Z x^B_X x^B_Z$,
where $x^I_i = N_i^I/V_I$$(i = X, Y, Z; I = A,B)$ and $N_i^I$ is the number of molecule species $i$.
The rates of replicating $Y$ and $Z$ are, respectively, given by 
\[ F_Y^A = V_A f_Y^A s_Y, \hspace{5mm} F_Z^B = V_B f_Z^B s_Z\]
Here, $f_Y^A = c_X x^A_X x^A_Y, f_Z^B = c_X x^B_X x^B_Z$.
$s_i = S_i/S_i^0$$(i=X, Y, Z)$ is the normalized concentration of the resource where $S_i$ is the concentration and $S_i^0$ is introduced to normalize $S_i$ to one when $S_i = S_i^0$.

The rate equations of cell-types $A$ and $B$ for molecule species $X$ are written as, 
\begin{equation}
\frac{dN_X^A}{dt} = F_X^A = V_A f_X^A s_X , \frac{dN_X^B}{dt} =  F_X^B = V_B f_X^B s_X, 
\label{RateX}
\end{equation}
and, for $Y$ and $Z$, 
\begin{equation}
\frac{dN_Y^A}{dt} = F_Y^A = V_A f_Y^A s_Y, \frac{dN_Z^B}{dt} =  F_Z^B = V_B f_Z^B s_Z. 
\label{RateYZ}
\end{equation}
The dynamics of resources $S_X$, $S_Y$, and $S_Z$ are respectively written as
\begin{align}
\frac{dS_X}{dt} &= - \left( V_A f_X^A + V_B f_X^B \right) s_X +DS_X^0(1 - s_X), 
\label{RateSX}
\\
\frac{dS_Y}{dt} &= - V_A f_Y^A s_Y  +DS_Y^0(1 - s_Y), 
\label{RateSY}
\\
\frac{dS_Z}{dt} &=  - V_B f_Z^B s_Z +DS_Z^0(1 - s_Z).
\label{RateSZ}
\end{align}

\begin{figure}[t]
\begin{center}
\rotatebox{270}{
\includegraphics[width=6cm]{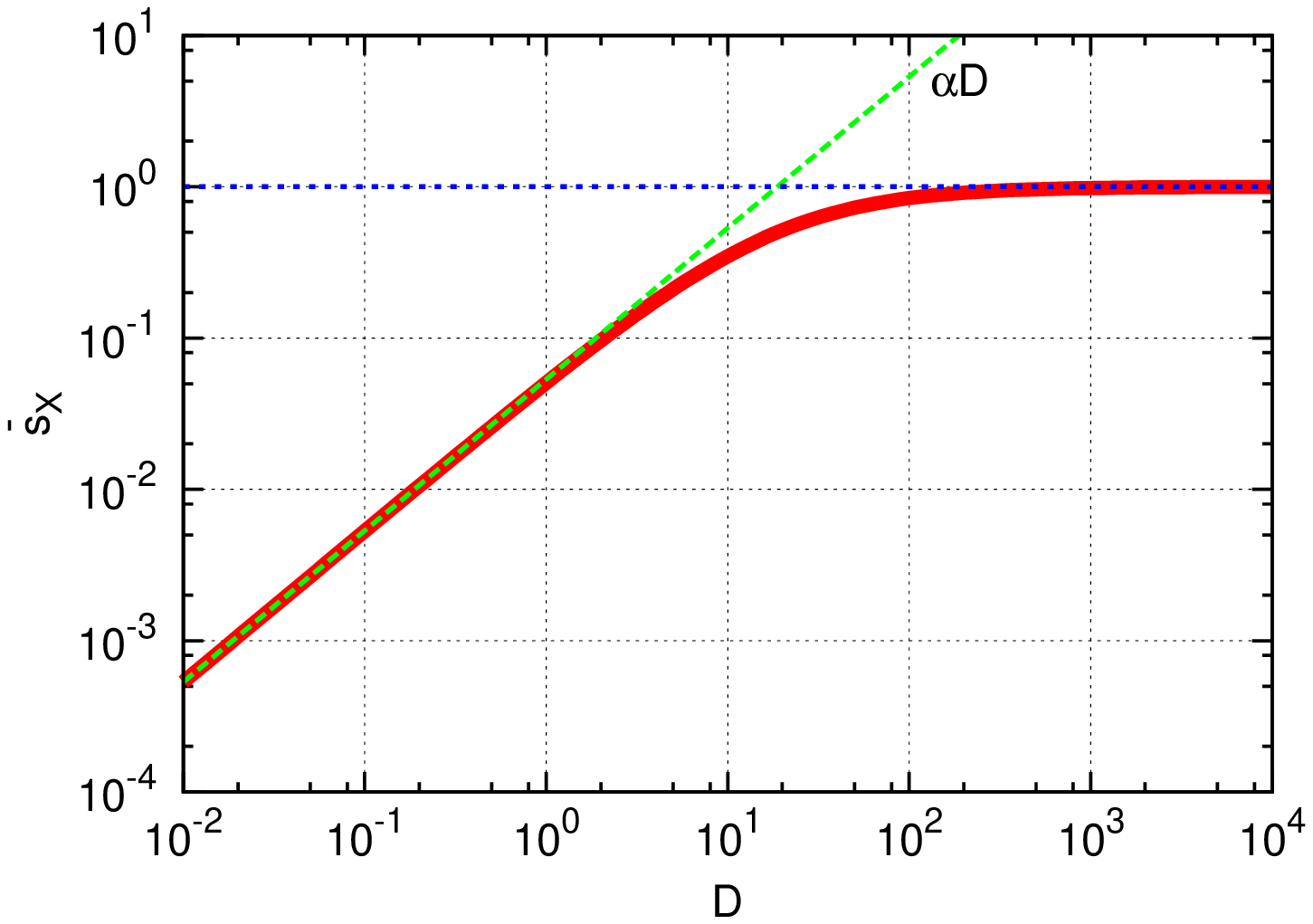}}
\vskip 0.25cm
\caption{The value of $\bar{s}_X$ as a function of $D$. The parameters are $c_Y = c_Z = 1$, and $S_X^0 = 1000$. The $x_i^I$ is fixed to $1/2$$(i=X, Y, Z, I = A, B)$, and $V_A = V_B = 750M_{\rm tot}/2$ with $M_{\rm tot} = 100$. The line $\alpha D$ is also shown with $\alpha = S_X^0/\left( V_A f_X^A + V_B f_X^B \right)$.}
\label{fig:0}
\end{center}
\end{figure}

In the steady state, the value of $s_X$, $\bar{s}_X$, is written as,
\begin{equation} 
\bar{s}_X = \frac{DS_X^0}{\left(V_A f_X^A + V_B f_X^B \right) + DS_X^0}.
\label{eq3}
\end{equation}
In Fig. \ref{fig:0}, we show $\bar{s}_X$ as a function of $D$ with a set of parameters. 
For large $D$, $\bar{s}_X \rightarrow 1$.
As $D$ is decreased, $\bar{s}_X$ starts to decrease and deviates from one. 
For smaller $D$, $\bar{s}_X$ decreases linearly as $\alpha D$. 

Similarly, for resources $S_Y$ and $S_Z$ the steady state values are written as
\begin{equation} 
\bar{s}_Y = \frac{DS_Y^0}{V_A f_Y^A + DS_Y^0}, \hspace{5mm} \bar{s}_Z = \frac{DS_Z^0}{V_B f_Z^B + DS_Z^0}.
\label{eq4}
\end{equation}

The dynamics of each molecule species changes with the diffusion constant $D$.
When $D$ is sufficiently large, $\bar{s}_X$ is approximately one. 
As $D$ decreases and the resource is limited, $\bar{s}_X$ deviates from one, and 
the maximum inflow rate of the resource is given by $J_X = DS_X^0(1 - \bar{s}_X)$.
In each range of $D$, we refer to $({\rm X-i}):$$S_X$ is sufficiently supplied, and $({\rm X-ii}):$ $S_X$ is limited.

For each range, the right-hand-sides of the rate equations (\ref{RateX}) can be written as 
\[ 
(F_X^A, F_X^B) = 
\begin{cases}
 (V_A f_X^A, V_B f_X^B)  \hspace{5mm} {\rm for } \hspace{1mm} ({\rm X-i})\\
 (r_A J_X, r_B J_X)  \hspace{5mm} {\rm for } \hspace{1mm} ({\rm X-ii})
\end{cases}
\]
where $r_A$ and $r_B = 1-r_A$ are the ratios of the resource being distributed to cell types $A$ and $B$. 

For the resources $S_Y$ and $S_Z$, we assume the range changes simultaneously for both of $Y$ and $Z$ because, otherwise, either type with a limited resource will vanish;  
we refer to ${\rm (YZ-i)}$: $S_Y$ and $S_Z$ are sufficiently supplied, and ${\rm (YZ-ii)}$: $S_Y$ and $S_Z$ are limited.
For each range, the right-hand-sides of the rate equations (\ref{RateYZ}) can be written as 
\[ F_Y^A =
\begin{cases}
    V_A f_Y^A \hspace{5mm}& {\rm (YZ-i)} \\
    J_Y \hspace{5mm}& {\rm (YZ-ii)}
  \end{cases}
  , F_Z^B =
\begin{cases}
    V_B f_Z^B \hspace{5mm}& {\rm (YZ-i)}\\
    J_Z \hspace{5mm}& {\rm (YZ-ii)}
  \end{cases}
 \] 
where $J_Y = DS_Y^0 (1-\bar{s}_Y)$ and $J_Z = DS_Z^0 (1-\bar{s}_Z)$.
As will be shown in subsequent subsections, we investigate four cases for the conditions $(X, YZ) = {\rm (i,i), (ii, i), (i, ii)}$ and ${\rm (ii,ii)}$.

The volume $V_A$, $V_B$ of each cell increases in proportional to the numbers of molecules as
\[ V_A = \frac{1}{R} (N_X^A+N_Y^A), V_B = \frac{1}{R} (N_X^B + N_Z^B), \]
where $R$ is a constant.

Then the increase in volumes is given by
\[ \frac{dV_A}{dt} = \frac{1}{R} (F_X^A + F_Y^A), \frac{dV_B}{dt} = \frac{1}{R} (F_X^B + F_Z^B),  \]

Let us suppose that a cell divides when its volume exceeds a certain critical value and that the total volume of all cells is restricted by some constant $\sum V_k = V_T = const$.
Then the volume fraction of each type $A$ and $B$ follows
\begin{align*}
\frac{dv_A}{dt} &= \frac{1}{RV_T} (F_X^A + F_Y^A) - \frac{v_A}{R} \sum \frac{F^I}{V_T}, \\
\frac{dv_B}{dt} &= \frac{1}{RV_T} (F_X^B + F_Z^B) - \frac{v_B}{R} \sum \frac{F^I}{V_T}.
\end{align*} 
where $v_I = V_I/V_T (I=A,B)$, $F_I = \sum F_i^I$.

\subsection{The case in which all the resources are sufficiently available: condition (X, YZ) = {\rm (i, i)}}
For the condition $(X, YZ) = {\rm(i, i)}$, when all resources are sufficiently available, the increases in molecule species are written as
\[ F_X^A = V_A f_X^A, F_Y^A = V_A f_Y^A, F_X^B = V_B f_X^B, F_Z^B = V_B f_X^B, \]
where 
\[ f_X^A = c_Y x^A_X x^A_Y, f_X^B = c_Z x^B_X x^B_Z, f_Y^A = c_X x^A_X x^A_Y, f_Z^B = c_X x^B_X x^B_Z, \]
and $x^I_i = N_i^I/V_I$$(I = A,B, i=X,Y, Z)$.
Thus, 
\begin{align*}
\frac{dv_A}{dt} &= \frac{f^A}{R}v_A - \frac{v_A}{R} \sum_I {v_I f^I}, \\
\frac{dv_B}{dt} &= \frac{f^B}{R}v_B - \frac{v_B}{R} \sum_I {v_I f^I},
\end{align*} 
where $f^A = f^A_X + f^A_Y, f^B = f^B_X + f^B_Z$.
The stationary states give
\begin{align*}
v_A f^A &= v_A \left( v_A f^A + v_B f^B \right), \\
v_B f^B &= v_B \left( v_A f^A + v_B f^B \right).
\end{align*}
The solutions to these equations are $(v_A, v_B) = (1,0)$ or $(0,1)$.

By writing $v_A = \delta$ and $v_B = 1-\delta$, linearization around $(v_A, v_B) = (0,1)$ gives
\begin{align*}
\frac{d\delta}{dt} &= \frac{f_A}{R} \delta - \frac{\delta}{R} (\delta f^A + (1-\delta) f^B) \\
&= \frac{\delta}{R} (f_A - f_B).
\end{align*}
Thus, the fixed point $(v_A, v_B) = (0,1)$ is stable when $f_B > f_A$. On the other hand, $(v_A, v_B) = (1,0)$ is stable when $f_A > f_B$.
Hence, the fittest cell type, i.e., that with larger $f_i$($i$=A or B) dominates the population.

\subsection{The case $S_X$ is limited: the condition (X, YZ) = {\rm (ii, i)}}
For the condition $(X, YZ) = {\rm (ii,i)}$, when $S_X$ is limited,
\[ F_X^A = r_A J_X, F_X^B = r_B J_X, F_Y^A = V_A f_Y^A, F_Z^B = V_B f_Z^B, \]
where $J_X$ is the maximum inflow of $S_X$ and $r_A$ and 
$r_B = (1-r_A)$ are the ratios of $S_X$ distributed into $A$ and $B$.
Then, 
\begin{align*}
\frac{dv_A}{dt} &= \frac{1}{R}\left( r_A j_X + f_Y^A v_A \right) - \frac{v_A}{R} \left( j_X + f_Y^A v_A + f_Z^B v_B\right) , \\
\frac{dv_B}{dt} &= \frac{1}{R}\left( r_B j_X + f_Z^B v_B \right) - \frac{v_B}{R} \left( j_X + f_Y^A v_A + f_Z^B v_B \right),
\end{align*}
where $j_X = J_X/V_T$.
The stationary condition is written as
\begin{align*} 
v_A \left\{ j_X + f_Y^A \left( v_A - 1\right) +f_Z^B v_B \right\} &= r_A j_X, \\
v_B \left\{ j_X + f_Z^B \left( v_B - 1\right) +f_Y^A v_A \right\} &= r_B j_X.
\end{align*}

The stationary condition has solutions $(v_A, v_B) = (1,0), (0,1)$.
The solution $(v_A, v_B) = (1,0)$ satisfies the condition with $r_A = 1$. Linearizing the rate equation around the fixed point
$(1,0)$ by writing $v_A = 1 - \delta$, we get 
\[ \frac{d\delta}{dt} = - \frac{1}{R} \left( f_Y^A - f_Z^B + j_X \right) \delta. \]
When $f_Y^A \geq f_Z^B$, the solution is stable. The solution $(v_A, v_B) = (0,1)$ satisfies the equation when $r_A = 0$. By linearizing the rate equation around the fixed point $(0,1)$ and writing $v_A = \delta$, we obtain
\[ \frac{d\delta}{dt} = - \frac{1}{R} \left( f_Z^A - f_Y^A + j_X \right) \delta. \]
When $f_Z^B \geq f_Y^A$, the solution is stable. 

For the case $f_Y^A = f_Z^B$, the equation is written as 
\[ \frac{d\delta}{dt} = - \frac{j_X}{R} \delta, \]
thus, both solutions are stable. 

\subsection{The case $S_Y$ and $S_Z$ are limited: condition $(X, YZ) = {\rm (i, ii)}$}
For the condition $(X, YZ) = {\rm (i,ii)}$ when $S_Y$ and $S_Z$ are limited, increases of molecule species are written as 
\[ F_X^A = V_A f_X^A, F_X^B = V_B f_X^B, F_Y^A = J_Y, F_Z^B = J_Z. \]
Then, 
\begin{align*}
\frac{dv_A}{dt} &= \frac{1}{R}\left( f_X^A v_A + j_Y \right) - \frac{v_A}{R} \left( f_X^A v_A + f_X^B v_B + j_Y + j_Z \right) , \\
\frac{dv_B}{dt} &= \frac{1}{R}\left( f_X^B v_B + j_Z \right) - \frac{v_B}{R} \left( f_X^A v_A + f_X^B v_B + j_Y + j_Z \right).
\end{align*}
The stationary state gives
\[
(f_X^A - f_X^B) v_A^2 + (-f_X^A + f_X^B + j_Y + j_Z) v_A - j_Y = 0.
\]

When $f_X^A = f_X^B$, this equation reduces to a solution
\[ v_A = v_A^0 = \frac{j_Y}{j_Y+j_Z}, v_B = v_B^0 = \frac{j_Z}{j_Y+j_Z}. \]
By linearizing the dynamics around the fixed point as $v_A = v_A^0 + \delta$, we get 
\[ \frac{d\delta}{dt} = -\frac{1}{R} \left( j_Y + j_Z \right) \delta, \]
and the solution is stable.

When $f_X^A > f_X^B$, the equation has a fixed-point solution with coexistence of $A$ and $B$ as 
\begin{align*} 
v_A = v_A^0 =  \frac{1}{2\left( f_X^A - f_X^B \right)}
\left\{ f_X^A - f_X^B - j_Y - j_Z 
+ \sqrt{(f_X^A - f_X^B - j_Y - j_Z)^2 + 4j_Y(f_X^A - f_X^B)}\right\}.
\end{align*}
As will be shown below, $0 < v_A^0 < 1$, so that the two types coexist: First, since $f_X^A - f_X^B > 0$, $v_A^0 > 0$.
Then the difference in the numerator and denominator in the expression of $v_A^0$ is given by
\begin{align*}
(numerator)-(denominator) = 
-(f_X^A-f_X^B + j_Y+j_Z) + \sqrt{(f_X^A - f_X^B - j_Y - j_Z)^2 + 4 j_Y (f_X^A - f_X^B)}. 
\end{align*}
Here, 
\begin{align*}
| f_X^A - f_X^B + j_Y + j_Z |^2 - \left\{ (f_X^A - f_X^B - j_Y - j_Z)^2+ 4 j_Y (f_X^A - f_X^B )\right\} 
= 4 j_Z (f^A_X-f^B_X) > 0,
\end{align*}
so that the denominator is greater than the numerator. Thus, $v_A^0 < 1$. Therefore, $0 < v_A^0 < 1$. 

By writing $v_A = v_A^0 + \delta$, and linearizing the dynamics around the fixed point, we obtain 
\[ \frac{d \delta}{dt} = - \left\{ (f_X^A-f_X^B) (2 v_A^0 - 1) + j_Y + j_Z \right\} \delta. \] 
The solution is stable if $(f_X^A - f_X^B)(2v_A^0 - 1) + j_Y + j_Z > 0$.
For example, if $f_X^A > f_X^B$ and $j_Y \geq j_Z$,the condition is satisfied.

When $f_X^B > f_X^A$, the stationary condition is
\[
(f_X^B - f_X^A)v_B^2 + ( -f_X^B + f_X^A +  j_Y + j_Z) v_B - j_Z = 0,
\]
for $v_B$. The fixed-point solution is obtained in the same way as in the case $f_X^A > f_X^B$, by replacing $A$ with $B$, and $Y$ with $Z$.

\subsection{All the $S_X$, $S_Y$ and $S_Z$ are limited: condition $(X, YZ) = {\rm (ii, ii)}$}
For the condition $(X, YZ) = {\rm (ii, ii)}$, when all resources are limited,
\[ F_X^A = r_A J_X, F_X^B = r_B J_X, F_Y^A = J_Y, F_Z^B = J_Z. \]
Then, 
\begin{align*}
\frac{dv_A}{dt} &= \frac{1}{R}\left( r_A j_X + j_Y \right) - \frac{v_A}{R} \left( j_X + j_Y + j_Z \right) , \\
\frac{dv_B}{dt} &= \frac{1}{R}\left( r_B j_X + j_Z \right) - \frac{v_B}{R} \left( j_X + j_Y + j_Z \right).
\end{align*}
The steady state gives
\[ v^0_A = \frac{r_A j_X + j_Y}{j_X + j_Y + j_Z}, v^0_B = \frac{r_B j_X + j_Z}{j_X + j_Y + j_Z}. \]
Again by writing $v_A = v_A^0 + \delta$,
\[ \frac{d \delta}{dt} = - \frac{j_X + j_Y + j_Z}{R} \delta, \]
follows, and the coexisting state is stable.

\subsection{Numerical simulations}

In this subsection, we present results of numerical simulations to show whether competition of limited resources results in dominance or coexistence of cells of type $A$ and $B$.

\subsubsection{Continuous simulation}
We have simulated the rate equations for the continuous concentration variables to confirm
coexistence by resource competition.

We consider $M_{\rm tot}$ cells of type $A$ or $B$.
The dynamics of the number $N_i^I$ of molecule species $i$($i = X, Y, Z$) in a cell-type $I$$(I=A,B)$ is written as eqs. (\ref{RateX}) and (\ref{RateYZ}). The dynamics of each resource $S_i$$(i=X, Y, Z)$ is given by eqs. (\ref{RateSX})-(\ref{RateSZ}).

As an initial distribution of cell types, types $A$ and $B$ are each represented at $50\%$.
In each type, the numbers $N_X$, $N_Y$ in type $A$, and $N_X$, $N_Z$ in type $B$ are randomly assigned. 
A cell divides when $V_K = \sum_i N_i^K$ exceeds a threshold $V_{\rm max}$.
At the division, the number of each molecule species $N_i^K$ is divided between the two daughter cells with a normal distribution with average $N_i^K/2$ and variances $N_i^K/4$. The normal distribution approximates the random partition of molecules into two daughter cells. 

\begin{figure}[t]
\begin{center}
\rotatebox{270}{
\includegraphics[width=6cm]{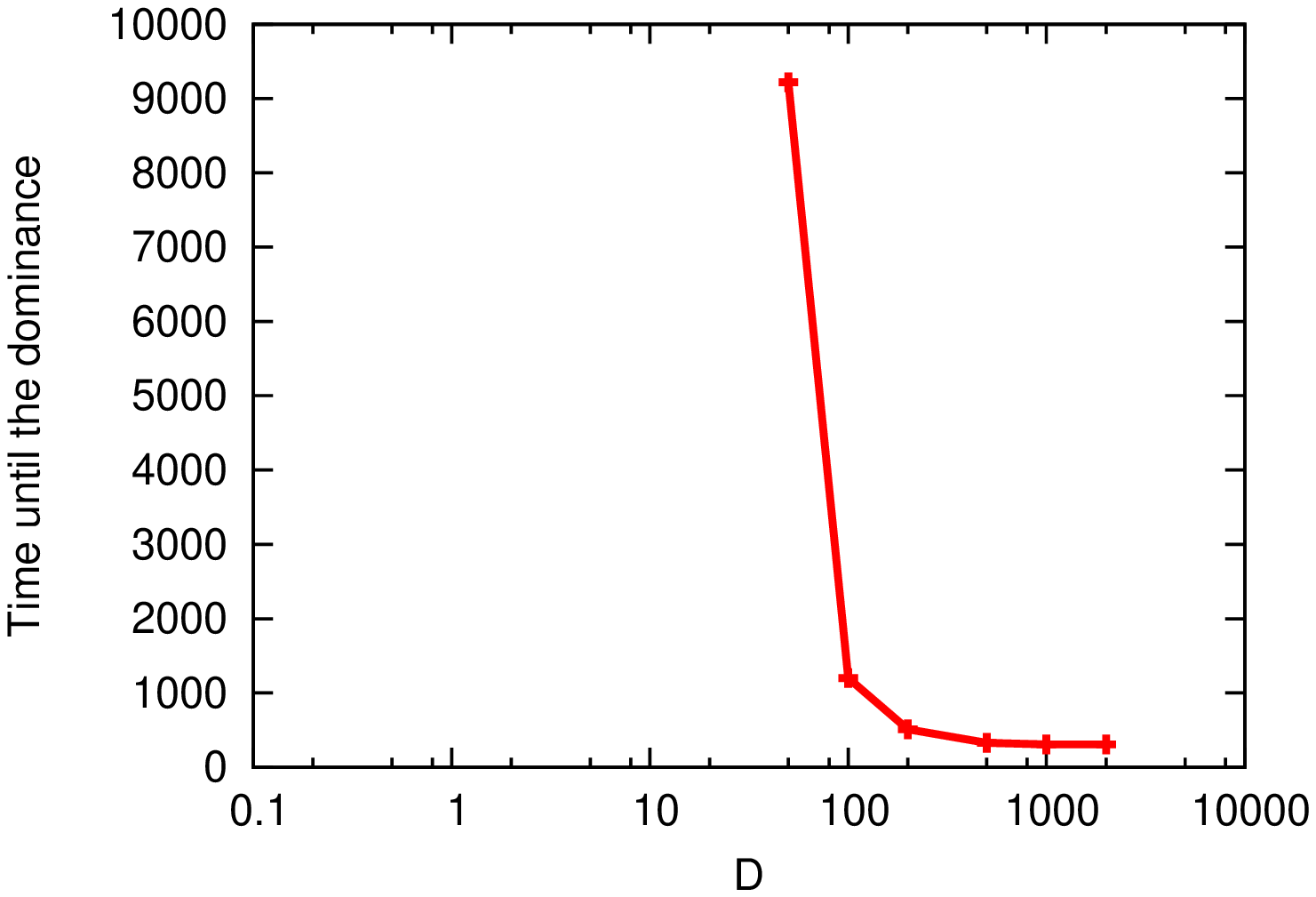}}
\vskip 0.25cm
\caption{Average time to a state in which either $A$ or $B$ dominates. Here, $M_{\rm tot}$ = 100, $V_{\rm max} = 1000$, $S^0_X = S^0_Y = S^0_Z = 1000$.}
\label{fig:three}
\end{center}
\end{figure}

\begin{figure}[t]
\begin{center}
\includegraphics[width=8cm]{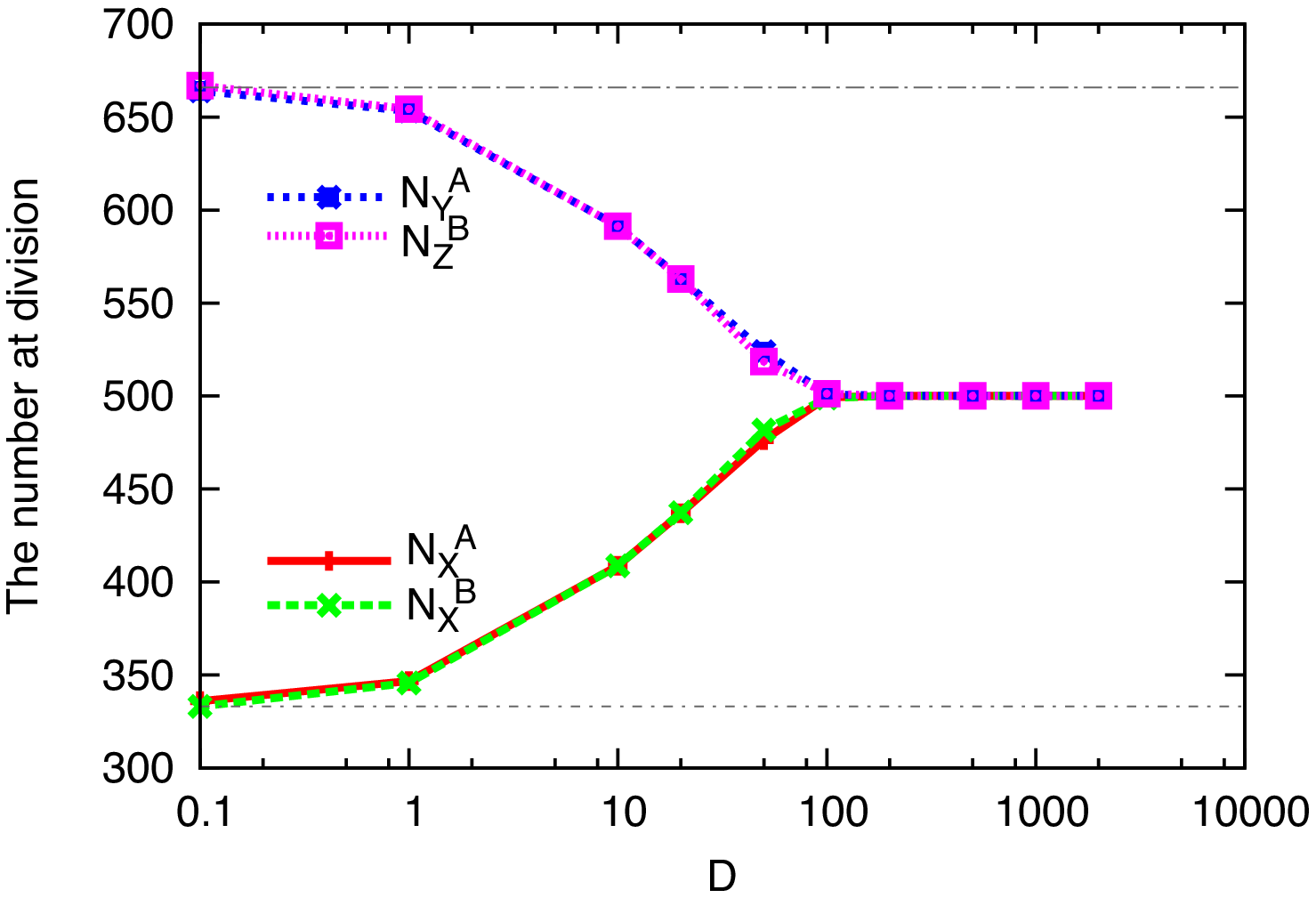}
\vskip 0.25cm
\caption{Average number of $N_X$, $N_Y$, and $N_Z$ at division events. The initial state is given as $n_A = n_B = M_{\rm tot}/2$, and in each type $A$ and $B$, we randomly distribute molecules $X$ and $Y$, and $X$ and $Z$, respectively, with equal probabilities. }
\label{fig:four}
\end{center}
\end{figure}

\begin{figure}[t]
\begin{center}
\rotatebox{270}{
\includegraphics[width=6cm]{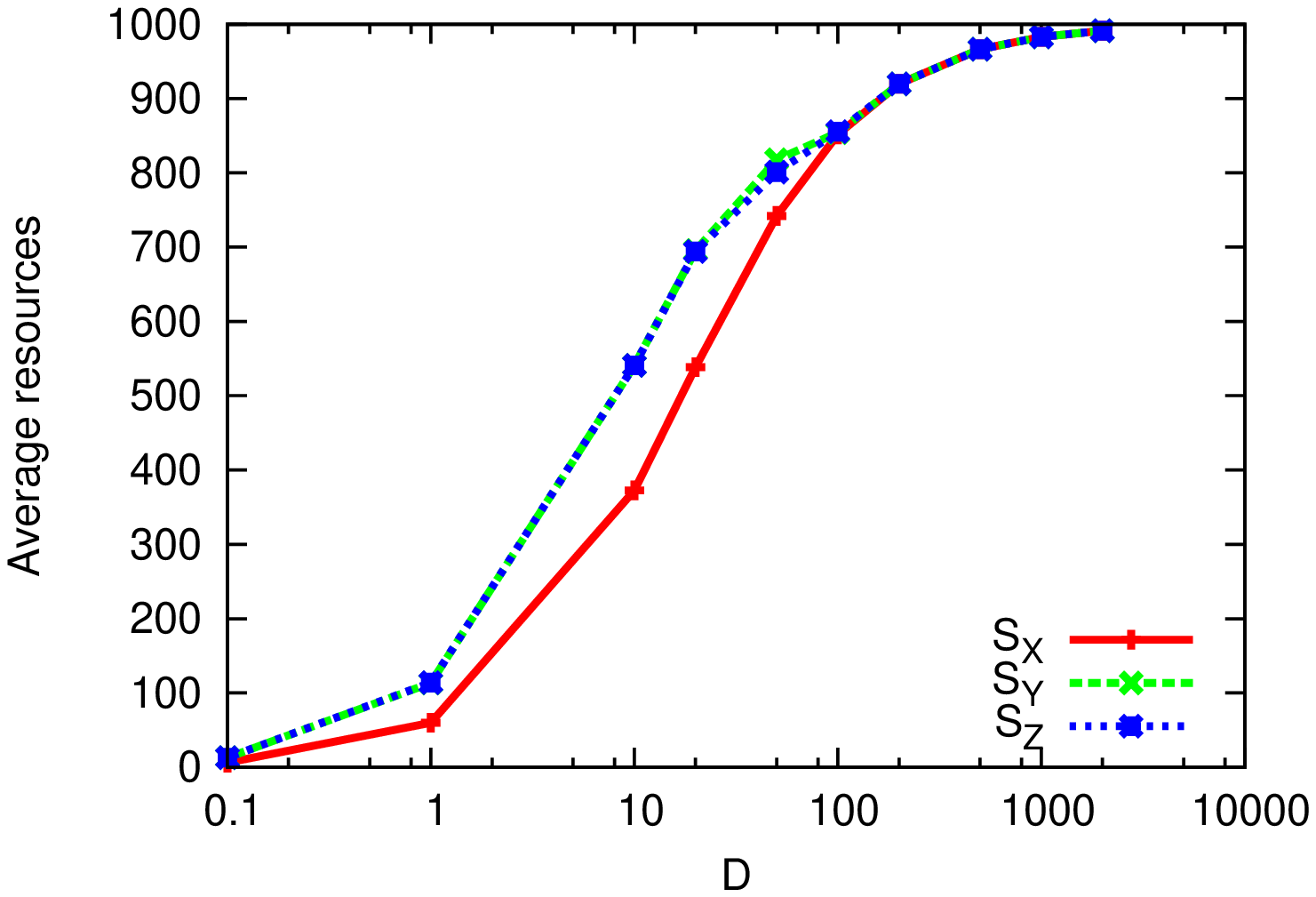}}
\vskip 0.25cm
\caption{Average resources $S_X$, $S_Y$ and $S_Z$. Here, $M_{\rm tot}$ = 100, $V_{\rm max} = 1000$, $S^0_X = S^0_Y = S^0_Z = 1000$.}
\label{fig:five}
\end{center}
\end{figure}

The average time required for the initial condition to reach a state in which either of the types is extinct is shown in Fig. \ref{fig:three}.
The parameters approximately correspond to those of Fig. \ref{fig:0}.
The time drastically increases around $D^* = 100$, which is consistent with the point where the resource $S_X$ starts to be limited in Fig. \ref{fig:0}.
Correspondingly, the average numbers of $N_X$, $N_Y$ and $N_Z$ at division events start to deviate below that point(Fig. \ref{fig:four}), and available resources show deviations between $S_X$ and $S_Y$ or $S_Z$ below that point(Fig.\ref{fig:five}).

At the point $D = D^*$, consumption and inflow of resources are balanced and $S_X$ starts to compete. 
The condition for balance in $S_X$ is 
\begin{align*}
\frac{M_{\rm tot} V}{4} \frac{S_X}{S_X^0} = D^* \left( S_X^0 - S_X \right),
\end{align*}
where $V$ denotes the average volume of cells.
By substituting $V = 750$, $S_X = 850$, $S_X^0 = 1000$, $M_{\rm tot} = 100$, $D^*$ is estimated as $\sim 100$.

For $S_Y$ and $S_Z$, competition starts roughly when the inflow is half that of $S_X$
because approximately half of $M_{\rm tot}$ cells consume the respective resource.
This suggests that below $D = 0.5 D^*$, all the resources are limited; thus, two types of cells coexist.
This is consistent with our numerical observations that below $D = 40$, the types coexist.

\subsubsection{Stochastic simulation}
We next show the results of stochastic simulations. The setup is given in the same way as in the main text. 
The resources $S_X$, $S_Y$, and $S_Z$ flow from the outer reservoir with diffusion constant $D$.
We consider $M_{\rm tot}$ cells, each of which consists of either pair of $X, Y$ or $X, Z$, corresponding to type $A$ or $B$, respectively.
For simplicity, we assume $c_X = c_Y = c_Z = 1$, $S^0_X = S^0_Y = S^0_Z =  M_{\rm tot}$. The diffusion constants are identical for $X$, $Y$ and $Z$.

\begin{figure}[t]
\begin{center}
\includegraphics[width=\textwidth]{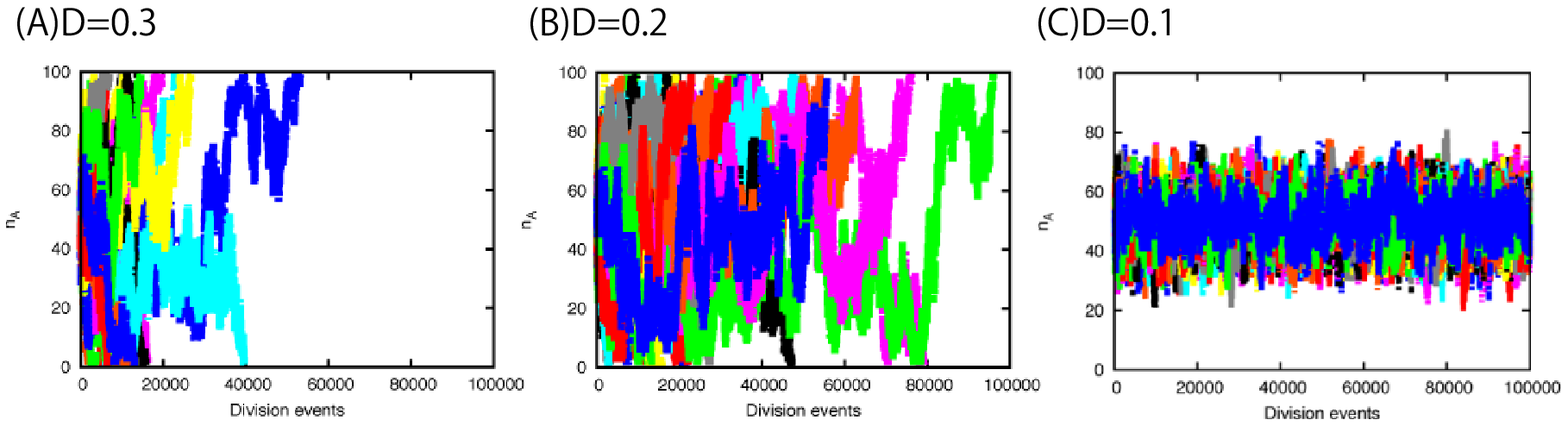}
\vskip 0.25cm
\caption{Time evolution of $n_A$ for (A)D = 0.3 (B)D = 0.2 and (C)D = 0.1. The initial state is given as $n_A = n_B = M_{\rm tot}/2$, and in each type $A$ and $B$, we randomly distribute molecules, $X$ and $Y$, and $X$ and $Z$, respectively, with equal probabilities.}
\label{fig:one}
\end{center}
\end{figure}

The evolution of the number of type $A$, $n_A$, for $D=0.3, 0.2$ and $0.1$ is 
given in Fig. \ref{fig:one}.
For $D=0.3$, either $A$ or $B$ is extinct after a relatively short time spam.
Around $D = 0.2$, the transient time before the extinction increases, and two types coexist over more than $10^5$ division events for $D=0.1$.

\begin{figure}[t]
\begin{center}
\rotatebox{270}{
\includegraphics[width=6cm]{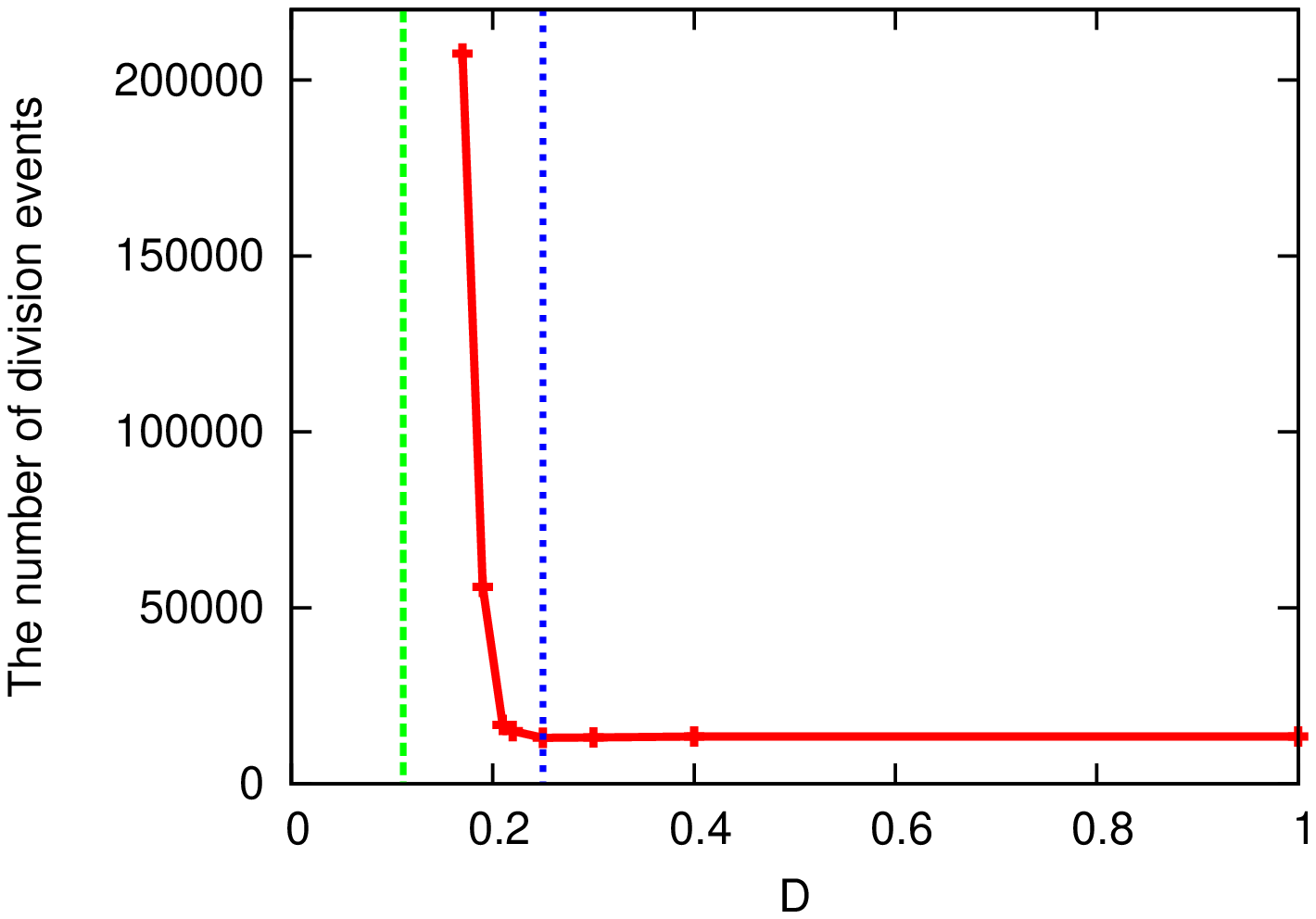}}
\vskip 0.25cm
\caption{Average number of division events, as a function of $D$, to achieve a state where either $A$ or $B$ cells dominates the system. Here, $M_{\rm tot}$ = $S_0$ = 100, $N = 1000$. The dotted lines show $D^* = 1/4$ and $D^+ = 1/9$.}
\label{fig:two}
\end{center}
\end{figure}

The average number of division events to achieve the dominant state as a function of $D$
is given in Fig. \ref{fig:two}. 
It is clear that the coexistence time starts to increase around $D^* = 0.25$, which is consistent with the point where the resources start to compete.
In the stationary state without competition, $N_X^A = N_Y^A$ and $N_X^B = N_Z^B$. 
Thus, the point where $S_X$ is limited is given by
\[D^* S_X^0 = \frac{1}{4} M_{\rm tot} \]

As $D$ is further decreased, the coexistence of types $A$ and $B$ is achieved when $S_Y$ and $S_Z$ are also limited and 
competed for by cells.
In this case, the coexistence is stable as analyzed in the previous subsection.
When all the resources are limited, the steady state approaches $N_X^A = 2 N_Y^A$, $N_X^B = 2N_Z^B$.
This gives
\[ D^+ S^0 \sim \frac{M_{\rm tot}}{2} \frac{2}{3} \frac{1}{3} = \frac{M_{\rm tot}}{9}. \] 
For $S^0 = M_{\rm tot}$, $D^+ \sim 0.11$. 
This is also consistent with the observation in Fig. \ref{fig:one}.

\newpage

\section{When different molecule species consume common resources : the case with $K_R < K_M$}
In this section, we investigate the case with $K_R < K_M$ where plural resource species are commonly used for
the replication of different molecules $X_i$.

\subsection{Model}
In the reaction
\[
X_j + X_i + S_{\hat{j}}  \rightarrow  2X_{j} + X_i,
\]
the correspondence between molecule $X_j$ and resource $S_{\hat{j}}$ is randomly assigned and fixed throughout simulations. 
Hence, each resource species is used commonly for $K_M/K_R$ reactions on the average.

\subsection{Results}

\begin{figure}[t]
\begin{center}
\includegraphics[width=16cm]{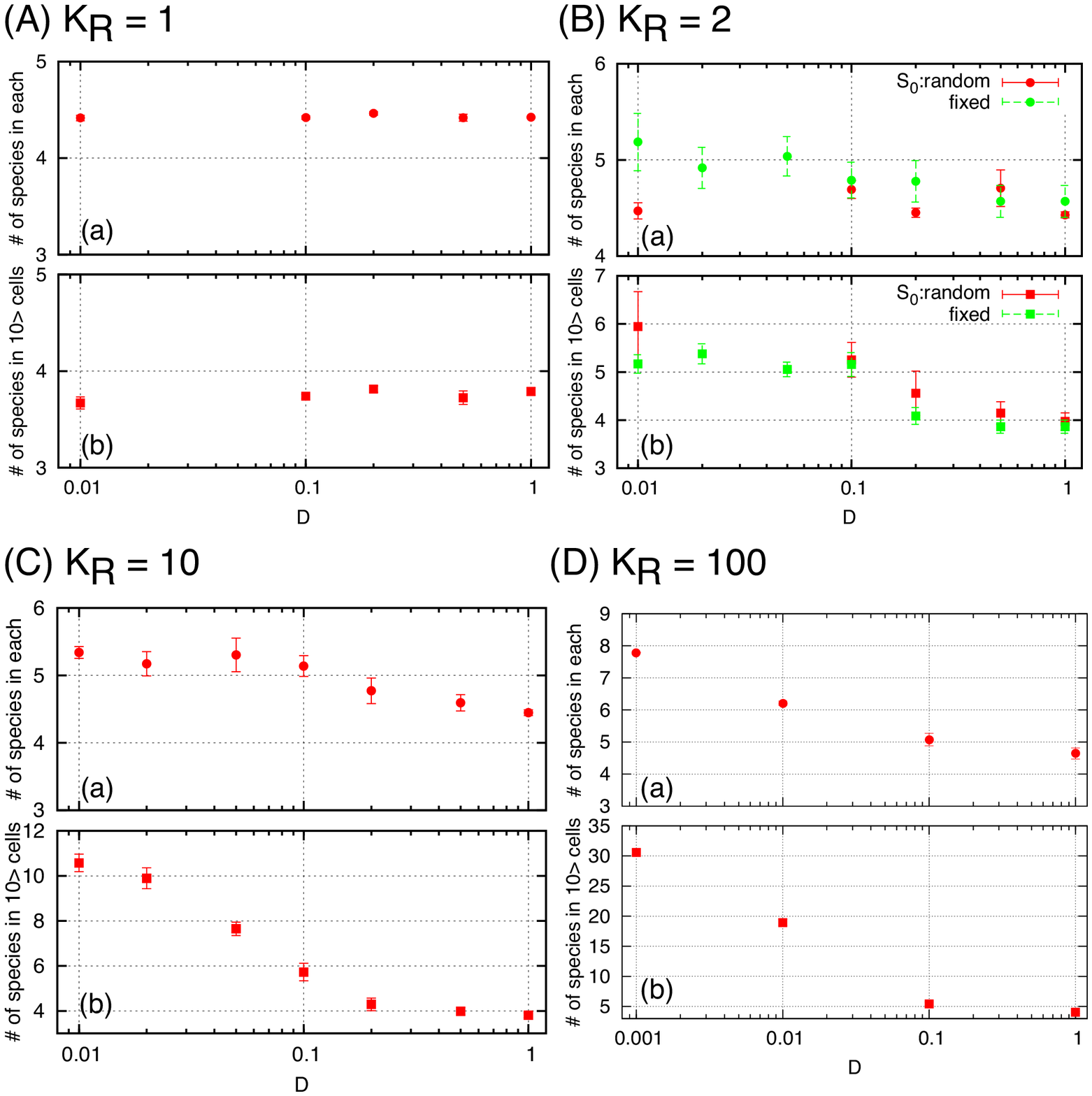}
\vskip 0.25cm
\caption{Compositional and phenotypic diversity plotted as a function of $D$ for (A) $K_R = 1$, (B) $K_R = 2$, (C) $K_R = 10$, and (D) $K_R = 100$. Compositional and phenotypic diversity are computed in the same way as in the main text, i.e., as the numbers of chemical species included in (a) each cell, and in (b) more than 10 cells out of $M_{\rm tot}$ cells are shown, respectively.
For (B)$K_R = 2$, the data for two cases are shown: $S_0^i$$(i=1,2)$ is determined randomly between 0 and $M_{\rm tot}$, and both $S_0^i$$(i=1,2)$ are fixed to $M_{\rm tot}$
All the data are obtained as the average over $10^5$ division events in 5 different networks with $M_{\rm tot} = 100$, $K_M =200$, $N = 1000$, and $\mu = 0.001$.}
\label{fig:six}
\end{center}
\end{figure}

We show the number of molecule species in each cells (compositional diversity) and that in 10 or more cells (phenotypic diversity), respectively, in Fig. \ref{fig:six}(A) for $K_R = 1$, i.e., a single resource is consumed to replicate all the molecule species $X_i$ $(i=1,...,K_M)$. In this case, neither compositional nor phenotypic diversity increases as the diffusion constant $D$ decreases. 

When two resource species are available($K_R=2$; Fig. \ref{fig:six}(B)), the phenotypic diversity increases as $D$  decreases, but a clear increase is not discernible in the compositional diversity. 
In the case of small $K_R$, the randomly determined reservoir concentrations, $S_0^i \in [0, M_{\rm tot}]$$(i=1,2)$, are also relevant parameters to determine the point at which the resources become limited, in addition to the diffusion constant.
Here, we also show the results for fixed $S_i^0 = M_{\rm tot}$$(i=1,2)$.
In contrast to the $K_M = K_R$ case, there is an increase in phenotypic diversity for $D < 0.1$, while even below the point, the number of remaining chemical species is approximately constant and does not increase as $D$ is decreased.
This result indicates that the diversity is bounded by the number of resource species: coexistence of at most two cell types is possible.

For larger $K_R$ ($K_R = 10$ in Fig. \ref{fig:six}(C), and $K_R = 100$ in Fig. \ref{fig:six}(D)), both compositional and phenotypic diversity increase as $D$ is decreased because $K_R$ is sufficiently large so that it does not effectively restrict the number of cell types. 
The phenotypic diversity, i.e., the number of coexisting cell types increases as $\sim K_R$, 
but the increase is saturated for larger $K_R$ (see Fig. \ref{fig:six}), which is also bounded by $M_{\rm tot}$. 
Indeed, as $M_{\rm tot}$ is increased, the number increases (see below).

\section{Dependence on $M_{\rm tot}$}
\begin{figure}[t]
\begin{center}
\rotatebox{270}{
\includegraphics[width=7cm]{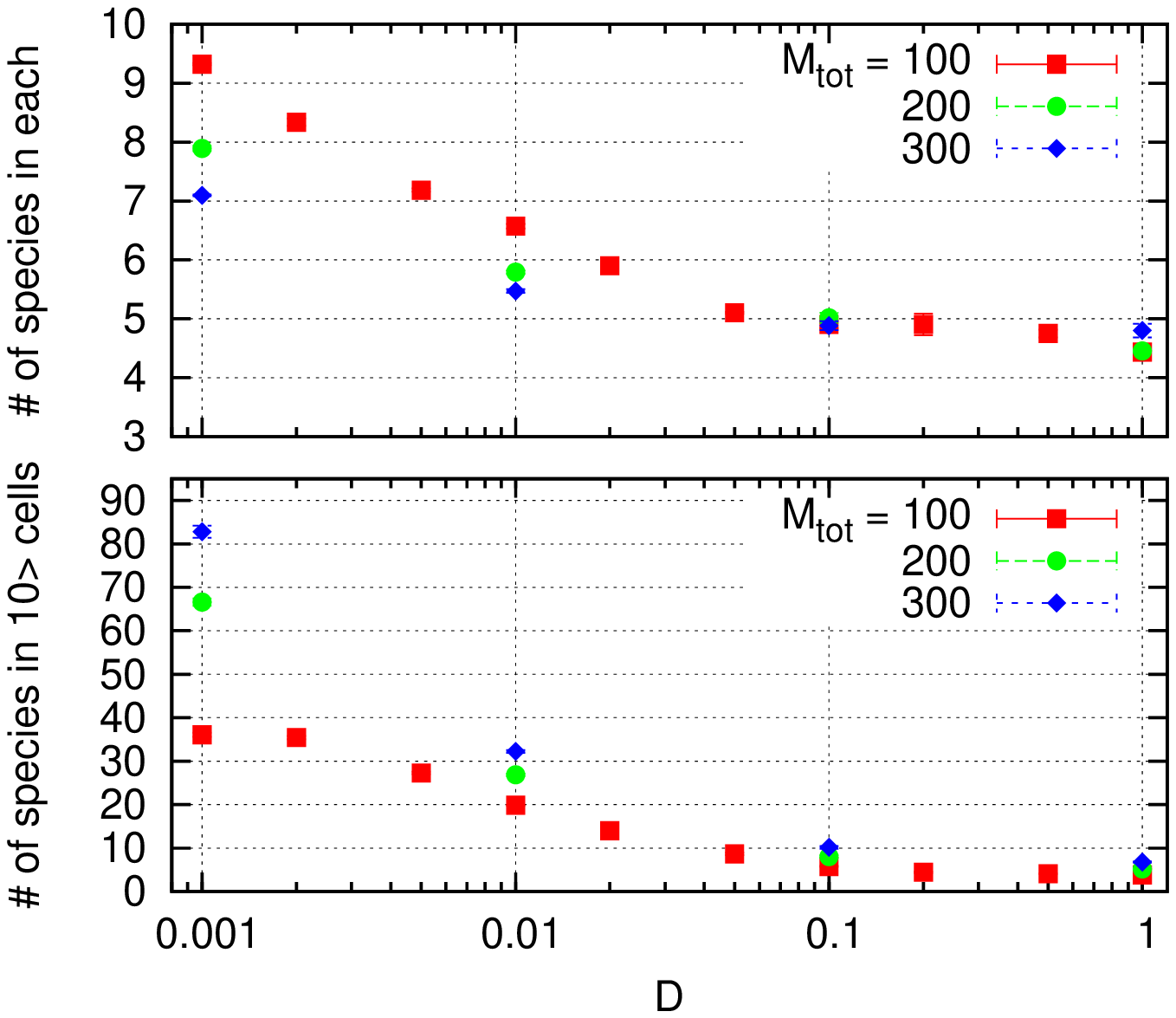}}
\vskip 0.25cm
\caption{Compositional and phenotypic diversity for $M_{\rm tot} = 100, 200,$ and $300$. 
The data are obtained as the average over $10^5$ division events in 30 different networks with $K_M = K_R = 200$, $N = 1000$, and $\mu = 0.001$}
\label{fig:ten}
\end{center}
\end{figure}

We investigated the dependence of the diversity on the number of interacting cells, $M_{\rm tot}$, in the case $K_M = K_R$. 
The number of molecule species in each cells (compositional diversity) and that in 10 or more cells (phenotypic diversity) for $M_{tot} = 100, 200$, and $300$ is shown in Fig. \ref{fig:ten}.

While both measures of diversity increase for each $M_{\rm tot}$, as $D$ is decreased, 
the increment depends on $M_{\rm tot}$. 
The increment in compositional diversity decreases as $M_{\rm tot}$ is increased.
On the other hand, the increment in phenotypic diversity 
with the decrease in $D$ increases as $M_{\rm tot}$ is increased. 
As shown in Fig. 3(C) of the main text, 
the two types of diversity clearly show opposite dependence on $M_{\rm tot}$ for fixed $D$. 

As discussed in the previous section, multiple cell types (up to $K_R$) can coexist; thus, 
more types of cells can be present and are not eliminated from the system as $M_{\rm tot}$ is increased, 
which results in the increase in phenotypic diversity.
On the other hand, as the number of cell types increases, 
a greater number of resource species are competitive because 
the cell population can consume more resource species. 
This leads each individual cell to be more specialized with fewer components, which results in suppression of the increase in compositional diversity.
 
\section{Supplementary figures}
\begin{figure}[t]
\begin{center}
\includegraphics[width=12cm, clip]{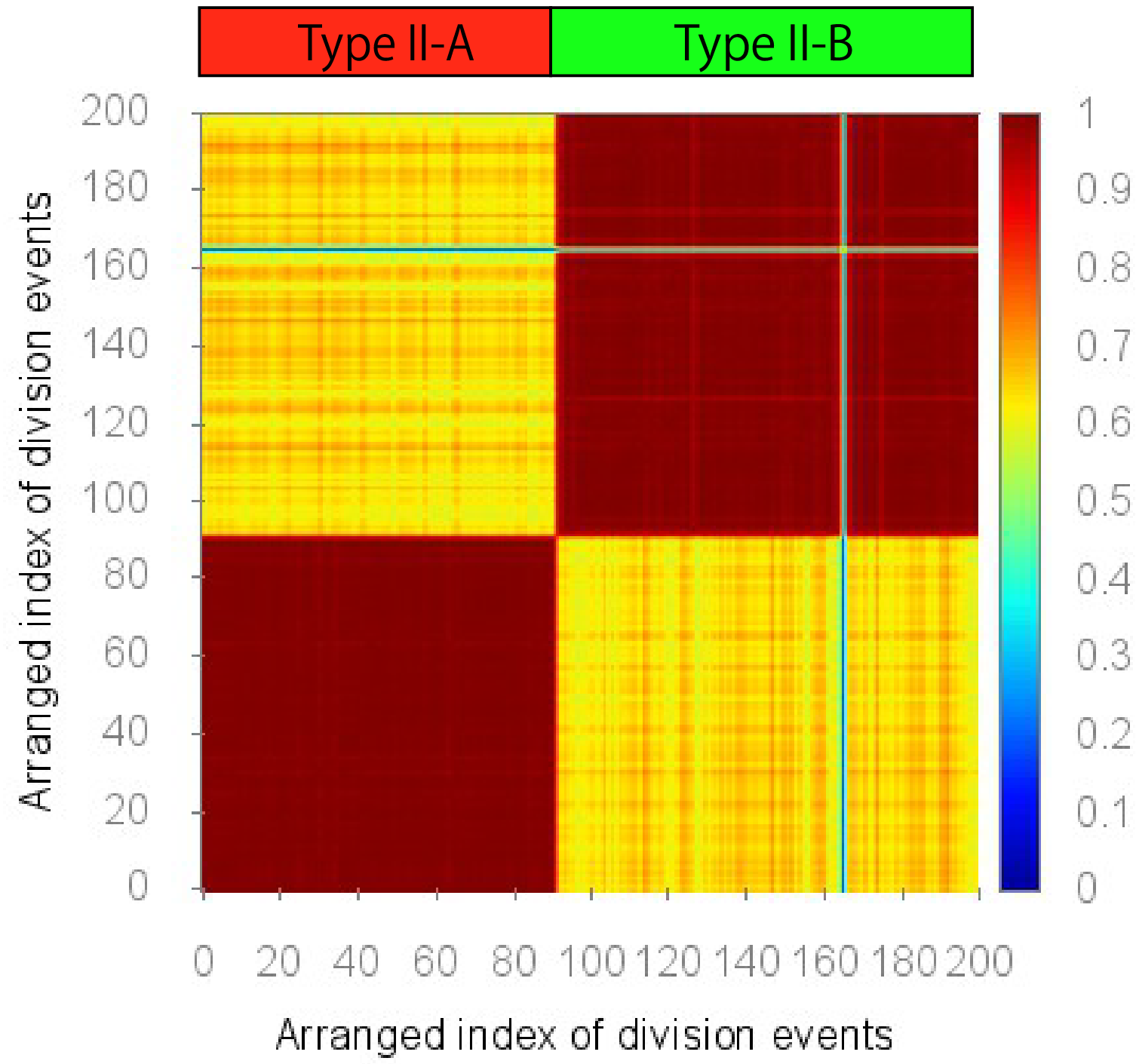}
\vskip 0.25cm
\caption{The indices are arranged from Fig. 2(II)(iii) in the main text to categorize into each type $A$ and $B$.}
\label{figS13}
\end{center}
\end{figure}

\begin{figure}[t]
\begin{center}
\includegraphics[width=12cm, clip]{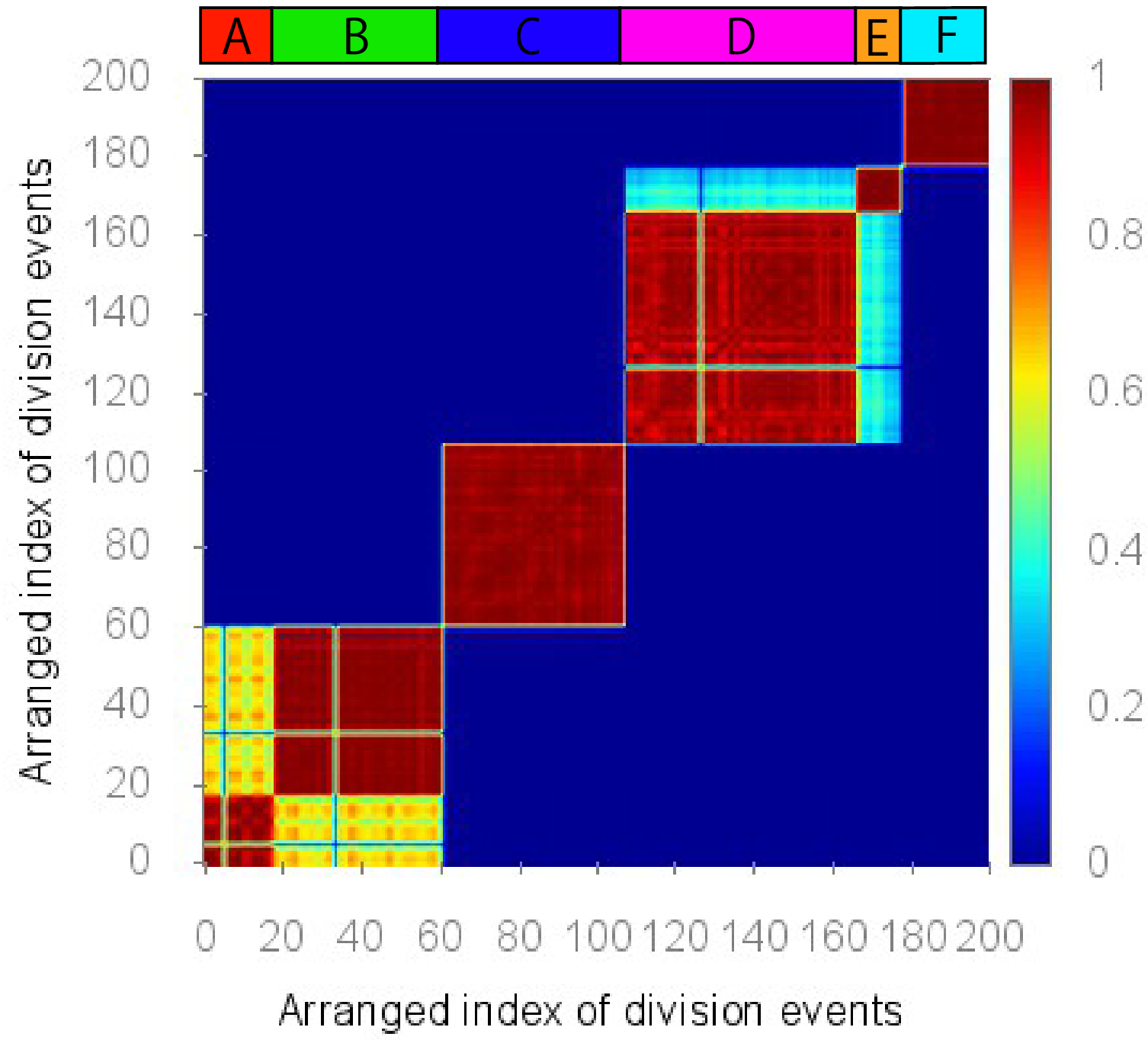}
\vskip 0.25cm
\caption{The indices are arranged from Fig. 2(III)(iii) in the main text to categorize into each type $A - F$.}
\label{figS14}
\end{center}
\end{figure}

Figs \ref{figS13} and \ref{figS14} show similarities $H_{ij}$ among a period of division events, where the cell indices given by the $x-$ and $y-$ axes are rearranged so that the same (similar) types are clustered. The data are identical to the similarities represented in Fig. 2(II)(iii) and (III)(iii) of the main text, but the plots are given after rearrangement of cell indices.

Types II-$A$ and II-$B$ are clustered as shown at the top of Fig. \ref{figS13}.
The similarities between cells of the same type have values close to 1, while types $A$ and $B$ have similarities around 0.6.

The types from III-$A$ to III-$F$ are clustered at the top of Fig. \ref{figS14}.
The similarities between cells of the same type have values close to 1, 
and those between types $A$ and $B$, and between $D$ and $E$ have positive values.
For other pairs, the values are almost orthogonal.

